\documentclass[draft]{agujournal2019}
\usepackage{url} 
\usepackage{lineno}
\usepackage[inline]{trackchanges} 
\usepackage{soul}
\draftfalse

\journalname{Journal of Advances in Modeling Earth Systems (JAMES)}
\usepackage{amsfonts}
\usepackage{gensymb}
\usepackage{algorithmic}
\usepackage{algorithm}
\begin{document}

\title{Ensemble methods for neural network-based weather forecasts}

\authors{S. Scher\affil{1} and G. Messori\affil{1,2}}

\affiliation{1}{Department of Meteorology and Bolin
Centre for Climate Research, Stockholm
University, Stockholm, Sweden}
\affiliation{2}{Department of Earth Sciences and Centre of Natural Hazards and Disaster Science (CNDS), Uppsala University, Uppsala, Sweden}

\correspondingauthor{Sebastian Scher}{sebastian.scher@misu.su.se}

\begin{keypoints}
\item We test four different methods to transform a deterministic neural network weather forecasting system into an ensemble forecasting system.
\item The ensemble mean of all methods is more skilful than a deterministic neural network forecast.
\item The spread-error correlation of the four methods is comparable to that of NWP forecasts.
\end{keypoints}

%
%

%
%


\begin{abstract}
Ensemble weather forecasts enable a measure of uncertainty to be attached to each forecast, by computing the ensemble's spread. However, generating an ensemble with a good spread-error relationship is far from trivial, and a wide range of approaches to achieve this have been explored -- chiefly in the context of numerical weather prediction models. Here, we aim to transform a deterministic neural network weather forecasting system into an ensemble forecasting system. We test four methods to generate the ensemble: random initial perturbations, retraining of the neural network, use of random dropout in the network, and the creation of initial perturbations with singular vector decomposition. The latter method is widely used in numerical weather prediction models, but is yet to be tested on neural networks. The ensemble mean forecasts obtained from these four approaches all beat the unperturbed neural network forecasts, with the retraining method yielding the highest improvement. However, the skill of the neural network forecasts is systematically lower than that of state-of-the-art numerical weather prediction models.
\end{abstract}

\section*{Plain Language Summary}
All weather forecasts are intrinsically uncertain. To address this issue, many modern weather forecasts rely on so-called ensembles. Rather than making a single ``deterministic" forecast (e.g. ``tomorrow will be sunny at location \textit{x}"), one performs a set (ensemble) of different forecasts. This set of forecasts represents different possible future weather scenarios (e.g. ``9 out of the 10 forecasts we performed show sunny weather tomorrow at location \textit{x}, while 1 shows cloudy skies"). Ensemble forecasts thus inform the user of the probabilities of certain weather outcomes, and also of the uncertainty of the prediction.
Recently, there has been growing interest in using a technique called neural networks for weather forecasting, instead of the conventional weather forecasting models. While conventional models start from the physical laws governing the atmosphere to perform weather forecasts, neural networks try to simulate the evolution of the atmosphere simply by "looking" at past observations. Here, we extend the neural network technique with methods borrowed from ensemble forecasting.

\section{Introduction}
For the last several decades, weather forecasting has been dominated by Numerical Weather Prediction (NWP) models, whose ongoing development has lead to a continuous increase in forecast skill \cite{bauer_quiet_2015}. Recently, there has been a growing interest in an alternative approach for weather prediction, through the use of neural-network based machine-learning
techniques. One use of machine-learning is to complement NWP forecasts. This can be done for example through using neural networks to correct NWP forecasts in a post-processing step, based on the errors of historical forecasts (e.g. \citeA{rasp_neural_2018-2}), to assign an \textit{ex-ante} confidence measure to forecast skill, also based on historical forecasts \cite{scher_predicting_2018-1} or to reduce the necessary number of ensemble members for spread estimation \cite{grnquist2019predicting}. Another approach is to train the neural networks on historical observations and then use them as a stand-alone tool to perform forecasts based on current observations. In this setting, they act as a prediction tool alternative to NWP models. Work in this direction has been done by \citeA{scher_toward_2018} and \citeA{scher_weather_2019-1} on simplified systems, and \citeA{dueben_challenges_2018,weyn_can_2019, weyn2020improving,rasp2020purely}
on reanalysis data. However, machine-learning forecast skill in the medium range ($\sim$ 3 -- 14 days) is typically much poorer than what operational NWP models achieve.

Independent of the forecast method, it has long been recognized that weather forecasts are more valuable when an uncertainty measure can be attached to them. This has led to the concept of probabilistic forecasting. While for point forecasts this can be achieved to some extent with statistical post-processing techniques \cite{glahn_use_1972-1}, the standard method to generate probabilistic forecasts is through the use of so-called ensemble forecasts \cite{ensNCEP1997,leutbecher_ensemble_2008}. In ensemble forecasting, NWP models are used to create several forecasts for the same time period, termed an ``ensemble''. The individual forecasts, namely the ensemble members, either have slightly different initial conditions, slightly different model formulations, stochastic components, or a combination of these. The development of ensemble forecasts at major weather forecasting centres began in the early 1990s, and was grounded in the realisation that the choice of initial perturbations is key to obtaining a skillful ensemble. Early studies in this field specifically highlighted the importance of identifying the fastest growing perturbations \cite{buizza1995optimal}. A detailed overview of the early development of initial perturbation techniques and ensemble forecasting is provided in \citeA{ensNCEP1997}. Other methods have been proposed to quantify forecast uncertainty, for example  measures derived from dynamical systems theory \cite{faranda_dynamical_2017-4}, and training a neural network on the error and spread of past NWP forecasts \cite{scher_predicting_2018-1}. However, ensemble forecasts remain the cornerstone of probabilistic weather forecasting. 

The simplest way of interpreting an ensemble forecast is to compute the spread of the ensemble members (here defined as the standard deviation of the ensemble members), and use this spread as a measure of confidence. In the ideal case, a high spread (namely a large difference between ensemble members) indicates high forecast uncertainty, while a low spread (a small difference between ensemble members) indicates a low forecast uncertainty. A further benefit of ensemble forecasts, beyond providing an estimate of forecast uncertainty, is that the mean of all members (the ensemble mean) has on average a higher forecast skill than when making a single (deterministic) forecast.

While there are a wide range of approaches to generate an ensemble, these may be grouped into two broad categories: 1) those that provide slightly different initial conditions for each member; and 2) those that in some way perturb/change the forecast model itself. Often, both categories of approaches are applied within the same ensemble forecasting system. The idea behind providing different initial conditions is to represent the uncertainty in our knowledge of the current atmospheric state. The creation of different initial conditions for the ensemble members, starting from the best guess that comes from a data-assimilation procedure
(e.g. \citeA{rabier_ecmwf_2000}), is a non-trivial task. The ECMWF's NWP model IFS -- usually recognised as delivering the world's best medium range weather forecasts -- uses a technique based on singular value decomposition (SVD) for this task \cite{molteni_ecmwf_1996}. Early versions used only the SVD technique, while newer versions use the SVD technique combined with an ensemble data assimilation scheme which itself already outputs different initial conditions. When applied in isolation, the latter results in too little spread among ensemble members (what is typically known as an ``underdispersive'' ensemble), and SVD is
therefore still in use \cite{ecmwf_ifs_2019}.  Perturbations of the model itself are usually performed within the parameterization schemes. For example, the physics perturbation scheme SPPT (Stochastically Perturbed Parametrisation Tendencies) of IFS randomly perturbs the tendency of certain atmospheric variables. The idea is that this represents the uncertainty in the approximations made in the parameterization schemes. This should, however, not be confused with random noise, as the perturbations are structured and applied at different spatial and temporal scales.

In this paper, we build upon the relatively recent development of neural networks as stand-alone weather forecast tools and extend them to incorporate probabilistic information. 
We specifically test four different methods for transforming a deterministic neural-network forecasting system into an ensemble forecasting system. Two of these perturb the initial conditions of the forecast, using random initial perturbations or an adaption of the SVD technique to neural networks. The two other methods perturb the forecast system itself, by retraining the neural network or randomly dropping weights in the network. SVD has already been proposed as part of a method to initialize the weights of neural networks \cite{bermeitinger_singular_2019}, but its use on trained networks for finding optimally perturbed input fields has, to our best knowledge, not been addressed before. 
We do not perform a comprehensive assessment of the probabilistic skill of the neural network forecasts, which would presumably be highly application-dependent. Rather, we focus on providing a proof-of-concept for performing neural network-based ensemble weather forecasts. Therefore, the methods described here are not designed to be competitive with the state-of-the-art NWP techniques.

Crucially, the word "ensemble" can be used differently in the contexts of machine learning and NWP models. When running an ensemble with a NWP model, the goal is to find  a set of possible future weather states. In the context of machine-learning, the term "ensemble" refers to all methods in which on or more algorithms are trained multiple times with slightly different settings, and the predictions averaged in order to get a better prediction, as for example in the widely used random forest algorithm. This is a much broader definition, as it could for example include the case where several models individually generate unrealistic forecasts, and only the mean of the forecasts is a skilful prediction. In this work, even though we use neural networks, we try to generate ensembles in the first sense, i.e. ensembles that represent a set of possible future weather states, even though we do not make the a-priori assumption that the only way to do this is through the modelling of growing instabilities as in NWP models. A second terminology issue concerns the term ``initial state''. In analogy to the NWP literature, we use the term ``initial state'' to refer to the input to the neural networks when forecasting. This is in contrast to the neural network literature, in which ``initial state'' sometimes refers to the initial weights of the neural network in the training procedure.

As discussed above, conventional NWP ensemble methods make us of both perturbations to the initial conditions and to the NWP model itself. These reflect two distinct sources of errors. Whether making this distinction explicit is also essential in the context of neural networks, is hard to answer a-priori. The neural networks do not intrinsically  attempt to provide a physically-grounded weather model, but rather are designed to optimise a specified output. Moreover, when training a machine-learning algorithm on (uncertain) atmospheric data, errors in the training data affect the algorithm's parameters. Assuming the machine-learning algorithm is trained and tested on different portions of a homogeneous dataset, this effectively conflates the error in initial conditions and error in the formulation of the machine-learning "model" itself. When performing a machine-learning forecast as we do in this study, the uncertainties in the initial state and construction of the system may therefore not be as distinct as in a conventional NWP context, although one could in principle separate the two. Here we therefore take a very applicative viewpoint, and focus on the question: ``do neural-network ensemble forecast have desirable statistical qualities when compared to the ground truth?", independent of how the forecasts are achieved. When evaluating and comparing our methods, we do not differentiate between perturbations of initial conditions and of the ``model" itself, and compare all approaches to one another. As a caveat, it may be argued that for hypothetical future neural-network forecast systems that compete in skill with NWP models, the distinction might become increasingly important. This underscores the more fundamental question of whether the distinction between initial condition and model errors is essential for any highly skilful forecasting system -- which we do not attempt to answer here. 

The four methods we adopt to generate neural-network forecasts are by no means the only possible approaches to implement probabilistic forecasts with machine-learning techniques. Other methods discussed in the literature include Bayesian neural networks and Generative Adversarial Networks (GANs). Bayesian neural networks are neural networks in which the weights of the networks are treated as random variables. Instead of learning a single value for each weight, in the training the distribution of the weight is learned (for example via mean and standard deviation). At prediction time, the value that is used for a particular weight is then drawn from this distribution. By making multiple predictions, an ensemble can thus be generated. An introduction to Bayesian neural networks is given in \citeA{jospin2020handson}. While Bayesian neural networks are an active field of research, to our best knowledge this technique has not yet been used in the context of weather forecasting. Another form of probabilistic neural networks are variational auto-encoders \cite{kingma2013autoencoding}, and the related GANs \cite{goodfellow2014generative}. GANs are used to infer high-dimensional probability distributions. In their conditional form, in which they infer a high-dimensional probability distribution conditioned on a (potentially also high-dimensional) input condition, they are very appealing for weather forecasting. At prediction time, an arbitrary number of samples can be drawn from the prediction distribution, thus resulting in an ensemble forecast. This has been demonstrated by \citeA{Bihlo2020AGA}, who were able to get skillfull 24-hour 2m temperature forecasts (but no skillfull precipitation forecasts). Further, they used a drop-out approach to generate ensembles of GANs. \citeA{GAgne2020} used GANs for stochastic parameterization in an idealized model. A further way of producing probabilistic forecasts with neural networks has been proposed by \citeA{sonderby2020metnet} for precipitation forecasting. The authors use a neural network that provides a discrete probabilistic output in the form of bins, and is able to outperform NWP forecasts for the first 7 to 8 hours.

In our study, we want to adopt a machine-learning approach already tested in the context of medium-range weather forecasting. Neural networks are amongst the machine-learning algorithms that have enjoyed the widest application as stand-alone weather forecasting tool. The four approaches we propose here to obtain ensemble forecasts are selected based on their applicability to neural networks. We specifically aim to assess the feasibility of using these four approaches to turn an existing neural network weather forecasting system into an ensemble forecasting system. As neural network system we use the architecture proposed by \cite{weyn_can_2019} and train it on 500hPa geopotential data from the ERA5 reanalysis. We then compare the four neural network ensemble approaches between themselves and with the results of the GEFS ensemble NWP model.

\section{Methods}

\subsection{Data}

We use atmospheric data from ERA5, which is ECMWF's most recent reanalysis product \cite{era5_qj}. A reanalysis provides the best guess of the past state of the global atmosphere on a 3d grid, by combining a forecast model with all available observations. We use 6-hourly data of 500hPa geopotential over the Northern Hemisphere on a 2.5\degree grid for the period 1976-2016 for training, and 2017-2018 for testing. This follows \citeA{weyn_can_2019} and \citeA{weatherbench}. Since the cost of computing the network Jacobians and subsequently the singular vectors is quite high (see below), in the testing period we only use one initial state from every second day (i.e. one state per 2 days). The data is normalized to zero mean and unit variance. As reference NWP forecast data, we use the 2nd version of the GEFS ensemble reforecasts \cite{hamill2013} from 2017-2018. The GEFS ensemble forecasts are a set of historical and operational NWP forecasts, all performed with the same model configuration. They consist of 10 perturbed members and an unperturbed control run. Here, we use forecasts at lead times of up to 5 days, initialized daily. As with the neural network forecasts, we consider only the Northern Hemisphere. We evaluate the GEFS forecasts at a 1\degree resolution (the standard resolution of the archived data). The analysis was repeated after regridding GEFS to the same resolution as the data used for training the neural networks, and the results were very similar (not shown).

\subsection{Neural network architecture}

Neural networks are a set of (nonlinear) functions with a -- potentially very large -- set of parameters. The parameters and functions are organized in layers. There is always an input and an output layer, and usually also one or more intermediate layers, termed "hidden layers". If there is more than one hidden layer in the network, then the network is a "deep" neural network. The parameters are fitted ("trained") on a certain target -- for example minimising the mean square error of a prediction. Before training the network, however, one has to select a network structure (the "architecture"). The choice usually results from a combination of intuition and testing. Here, we rely on a network architecture previously tested in the literature for reanalysis climate data. Specifically, we adopt the purely feed-forward architecture presented in \citeA{weyn_can_2019}, with 500hPa geopotential as input. Feed-forward means that the network (and each individual layer) has an input and an output side, and the output is not redirected to the input. This is one of the most widely used neural network types. The \citeauthor{weyn_can_2019} architecture was developed for a different reanalysis product to the one we use here, namely the Climate Forecast System (CFS) Reanalysis. We regridded ERA5 to the same horizontal resolution as in \citeauthor{weyn_can_2019}, namely 2.5\degree, and we assume that the architecture is not overly sensitive to the change of reanalysis product.

\begin{figure}
\centering
\includegraphics[width=0.5\textwidth]{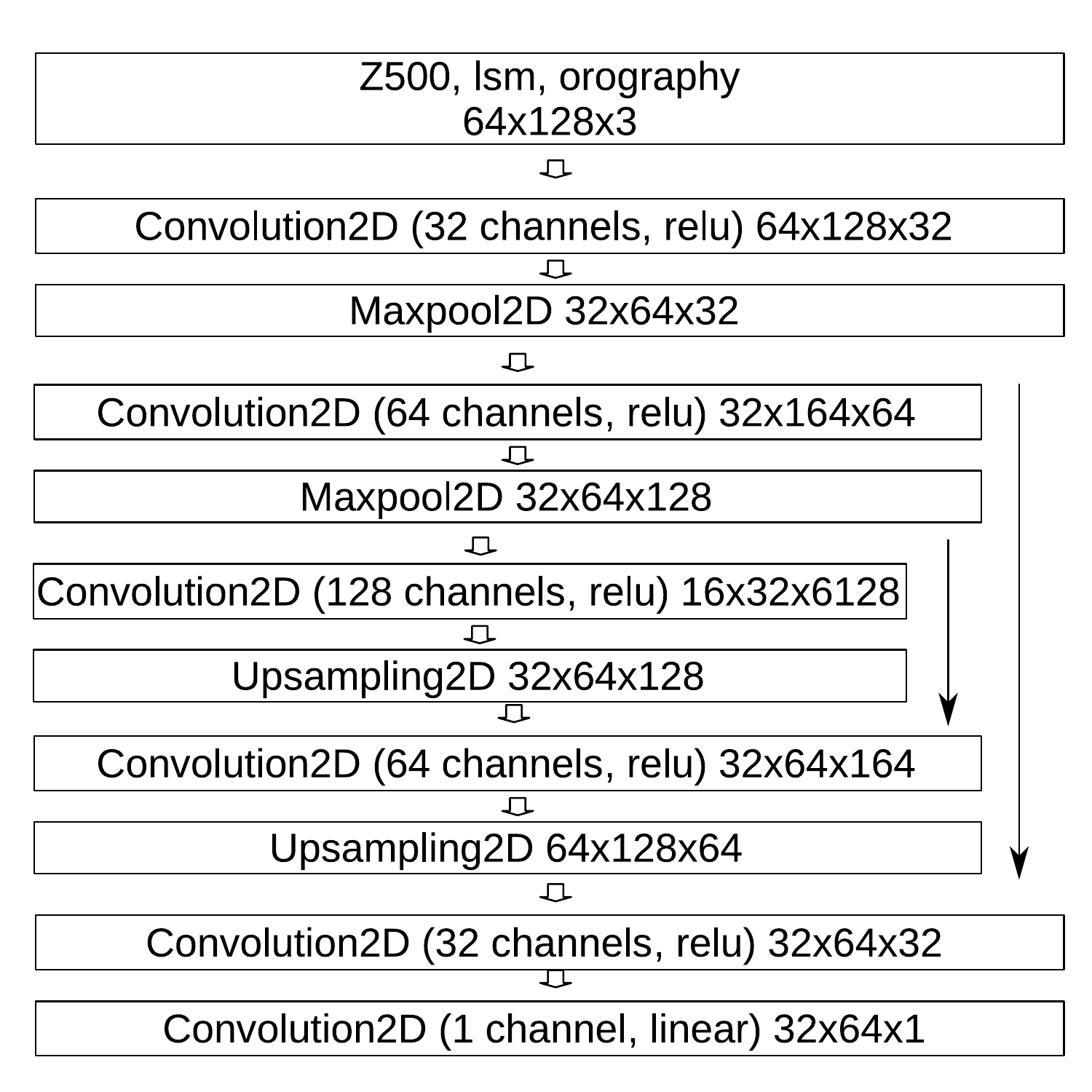}
\caption{\label{fig:network}The architecture of the neural network used in this study, based on \citeauthor{weyn_can_2019}.}
\end{figure}

The networks are trained with the Adam optimizer \cite{kingma_adam_2017} and mean square error loss. They are trained first for 200 epochs (iterations through the training data), and then additionally as long as the loss on the validation data (last 2 years of the training data) does not decrease anymore, with a maximum of 200 additional epochs. The networks are trained to forecast one
timestep (6 hours). Longer lead-times are obtained through consecutive one-step forecasts. Details of the network architecture are shown in fig. \ref{fig:network}.

\subsection{Ensemble techniques}
In this section, we describe the four different methods which we use to create ensembles of neural networks. Each of the methods uses the same neural network architecture. 
With each of the ensemble methods, 100 member are generated, except for the retraining ensemble, for which the maximum ensemble size is 20 due to constraints in available computation time.

\subsubsection{Random initial perturbations}
One of the conceptually simplest -- even though not necessarily best -- ways of creating an ensemble in a chaotic dynamical system is to perturb the input initial conditions with random noise. This can be done for any type of numerical model that accepts initial conditions, and is equally applicable to a neural network forecast that is initialised from an input vector (its ``initial condition''). In a conventional NWP model, one should place particular care in how these initial perturbations are propagated. In general, a naïve implementation of random perturbations is not an effective approach to generate an ensemble \cite{du2018ensemble}, and encouraging results for simple systems may not be representative of applications to the real atmosphere \cite{bowler_comparison_2006}. Perhaps surprisingly, we find that this simple approach seems to be relatively well-suited to our neural network forecasts (Sect. 3, 4).

We implement random perturbations following the method from \citeA{bowler_comparison_2006}, with the simplification that we use a pure Gaussian distribution, instead of the convolution of a Gaussian and an exponential distribution. This amounts to adding a value from a Gaussian distribution with zero mean and standard-deviation $\sigma_{rand}$, independently to each gridpoint and each ensemble member. Since the ensemble has finite size, the mean and standard deviation of the perturbations over the whole ensemble do not necessarily match those of the parent distribution. Therefore, the drawn samples are first normalized to zero mean and $\sigma_{rand}$ standard deviation. The variable $\sigma_{rand}$ is a free parameter that is varied experimentally (0.001, 0.003, 0.01, 0.03, 0.1, 0.3, 1, 3). We hereafter refer to this method as ``random ensemble'' (``rand'' in plot legends)

\subsubsection{Singular value decomposition}
Singular Value Decomposition (SVD) is a technique from linear algebra with a wide range of applications. One of its uses is to find optimal perturbation patterns for a function

\begin{equation}
f:\vec{x}
\in
\mathbb{R}^{N}
\rightarrow
\vec{y}
\in
\mathbb{R}{}^{M}
\end{equation}

where optimal means that the (infinitesimal) input perturbation pattern \textrm{$\vec{x}'$} leads to the maximum output perturbation \textrm{$\vec{y}'$} with respect to some norm, when linearizing the function around its input:

\begin{equation}
max\frac{||\vec{y}'||}{||\vec{x}'||}
\end{equation}

For example, if one imagines a simple system which has temperature as its only variable, this would be equivalent to looking for the (infinitesimal) perturbation in the input field that maximizes the change in prediction with respect to the unperturbed case.
In this paper, $N=M$ (input and output dimension are the same) and we use the euclidean norm, which allows the use of standard SVD routines from numerical libraries.

To find the singular vectors, we first compute the Jacobian of the
neural network:

\begin{equation}
J=\left[\begin{array}{cccc}
\frac{\partial y_{1}}{\partial x_{1}} & \frac{\partial y_{1}}{\partial x_{2}} & \cdots & \frac{\partial y_{1}}{\partial x_{M}}\\
\frac{\partial y_{2}}{\partial x_{1}} & \frac{\partial y_{2}}{\partial x_{2}} & \cdots & \frac{\partial y_{2}}{\partial x_{M}}\\
\vdots & \vdots & \ddots & \vdots\\
\frac{\partial y_{N}}{\partial x_{1}} & \frac{\partial y_{N}}{\partial x_{2}} & \cdots & \frac{\partial y_{N}}{\partial x_{M}}
\end{array}\right]
\end{equation}

This is simple to implement, as gradients of the output of neural
networks are central in the training procedures, and therefore neural network libraries usually provide functions to compute gradients and Jacobians of the output with respect to the network inputs. If no explicit function for Jacobians is available, then one can use the gradient function to compute all rows of the Jacobian individually through looping over the output dimension. We use the Jacobian functions of Tensorflow. Computationally, however, computing the Jacobian is relatively expensive, as it requires one gradient computation for each element of the output space. We use a lead-time of $T_{svd}$ hours for the computations of the input perturbation patterns. This means that, for a given
input $\vec{x}$, the Jacobian matrix of the function defined by the corresponding number of consecutive neural network forecasts is computed. We then compute the $n_{svs}$ leading singular vectors $S_{i}$ of the Jacobian matrix with a standard SVD-routine from the numpy library. This results in the SVD of the system, using the euclidean norm. The validity of the Jacobian is tested with a simple TLM test (Appendix B).

Following \citeA{bowler_comparison_2006}, the leading singular
vectors are then combined with random weights $a_{i}$ from a truncated Gaussian distribution with standard deviation $\sigma_{svd}$, truncated at $3\cdot \sigma_{svd}$ (since the random and the SVD ensemble technique do not necessarily use the same scales, we give them separate names).

\begin{equation}
\vec{x}_{pert}=\vec{x}\pm\sum a_{i}S_{i}
\end{equation}

This creates pairs of symmetric perturbations centered at zero (always one member with $+,$ one member with $-$). The algorithm
is sketched in fig. \ref{fig:method} and presented in detail in alg. \ref{alg:computation-of-the} in Appendix A. 

The parameters  $\sigma_{svd}$, $T_{svd}$ and $n_{svs}$ are varied experimentally (tested values: $\sigma_{svd}$: $[0.001,0.003,0.01, 0.03,0.1,0.3,1,3]$, $T_{svd}$: $[1,2,4,8]$ steps, corresponding to $[6,12,24,48]$ hours, and $n_{svs}$: $[10, 20, 40, 60, 80, 100]$). All cross-combinations are tested.
We hereafter refer to this method as SVD ensemble (``svd''
in plot legends). Due to the relatively high expense of computing the full Jacobians, only every second
day was used as initial state. Therefore, in order to allow for a
valid comparison, also the ensembles generated with the other methods were initialized every second day.

\begin{figure}
\includegraphics[width=0.8\textwidth]{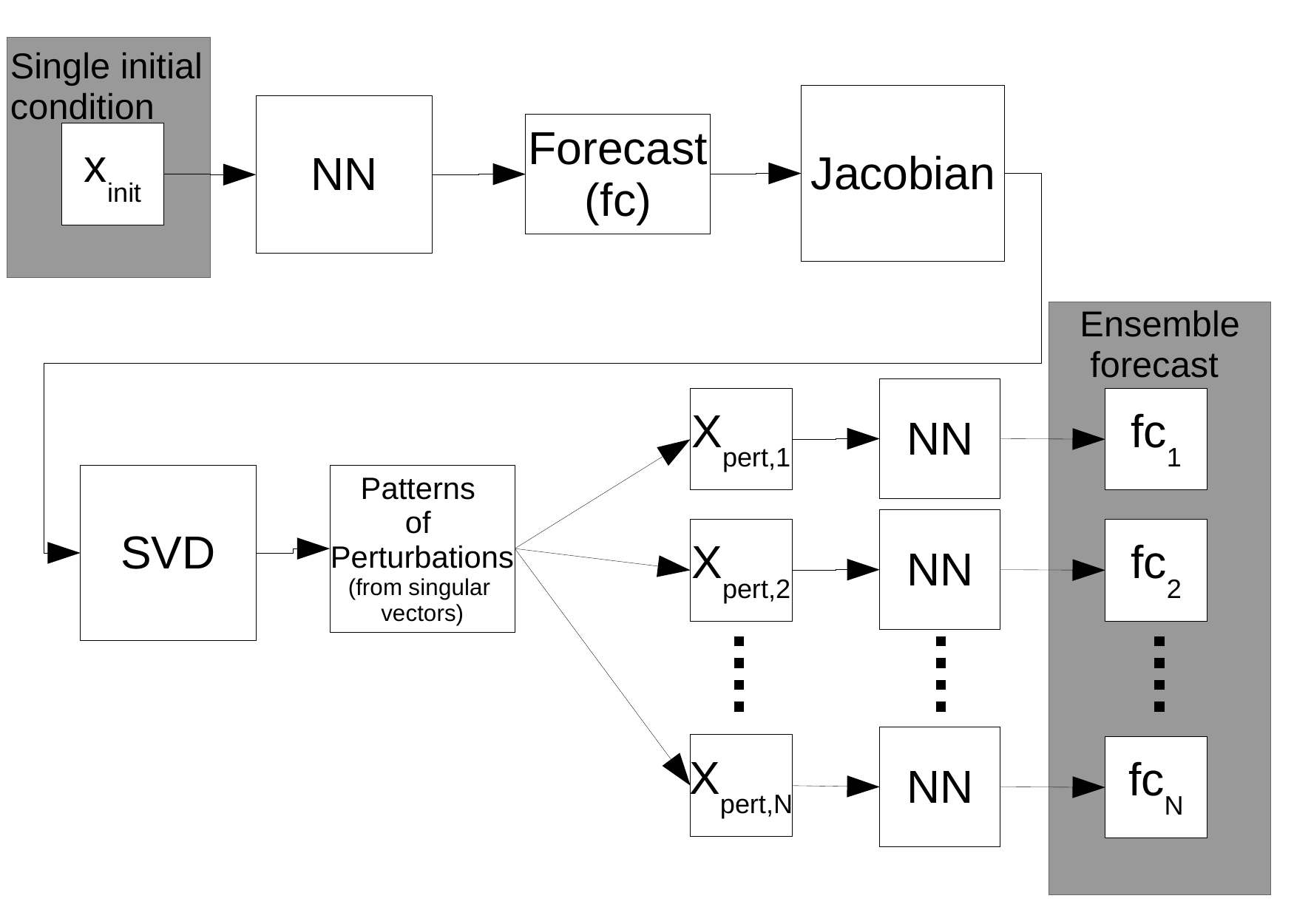}\caption{\label{fig:method}Overview of the SVD technique for creating ensemble
forecasts with neural networks (NN).}

\end{figure}

\subsubsection{Network retraining ensemble}

The neural network training procedure used here has two random components, namely the random initialization of the network weights, and the random selection of training samples in the training loop of the optimizer. Therefore, a simple way to create an ensemble is via retraining the network, starting with different initial seeds for the random number generators. It would also be possible to add another level of randomness via selecting a different subset of the training data for each member, although we have not explored this possibility here.

There is large variation in skill between different training realizations (some training realizations provide very poor forecasts at longer lead times, even though they have small errors on the training lead time of 6h). Therefore,  we trained 50 members, and then selected the 20 members that had the highest skill on the last year of the training data at a lead time of 60h. A leadtime of 60h was chosen because it is in the range where the network forecasts have reasonable skill. This ensemble will be referred to as the “\emph{multitrain} ensemble” (“multitrain” in plot legends)

\subsubsection{Dropout ensemble}

Dropout is a widely used regularization technique (that is, a technique to reduce overfitting) for neural networks \cite{JMLR:v15:srivastava14a}. When using dropout as a regularization method during training, for each iteration through the network a random selections of neurons (and their connections) are deactivated. Thus, only part of the network is trained in each iteration. When using the networks for predictions, no dropout is usually implemented, such that the whole network is used, resulting in a deterministic prediction. Here we use the dropout technique in an unconventional fashion. Instead of applying it during the training as regularization technique, or both during training and forecasting, we apply it only when using the trained network to make our forecasts. In other words, we first train the network without dropout, and then add the dropout to the trained network. After each convolution layer (except for the final linear one), we insert a dropout layer, which when used for predictions has a dropout probability of $p_{drop}$. Thus, for each forecast, the fraction $p_{drop}$ of the neurons in each layer is deactivated. The following values of $p_{drop}$ were tested: $[0.001,0.003,0.01,0.03,0.1,0.2,0.4]$   This ensemble will be referred to as the “\emph{drop} ensemble” (“drop” in plot legends).

\subsubsection{Unperturbed reference forecasts}

Similar to the unperturbed "control" runs of NWP ensembles, we use the individual members of the \emph{multitrain} ensemble as unperturbed forecasts. Using a single member is equivalent to training the network only once. To account for the randomness in the training and its potential influence on the skill of unperturbed forecasts, we compute the error for each member individually, and then average over all members to obtain a representative score. This will be referred to as "unperturbed".

\subsection{Ensemble spread, error and skill}

In a perfect ensemble, the spread of a forecast is the expectation
value of the error of the forecast, where spread is defined as standard deviation of the members, and error as the Root Mean Square Error (RMSE) of the ensemble mean \cite{palmer2006ensemble}. Thus, for a perfect ensemble, the mean
forecast spread should be equal to the mean forecast error, when averaged over many forecasts. Since standard deviations and RMSE cannot easily be averaged (a common pitfall when evaluating ensemble forecasts, see \citeA{fortin_why_2014}), we first compute the variance of the members at each gridpoint, and then average over all gridpoints. For mean spreads over multiple forecasts, we average over forecasts as well, and then finally take the square root to get the mean standard deviation. We follow the same procedure for the RMSE of the forecasts,
for which we compute the Mean Square Error (MSE), then average, and then take the square-root.

\begin{equation}
spread=\sqrt{\overline{var\left(\vec{y}_{ens}\right)}}
\end{equation}

\begin{equation}
RMSE=\sqrt{\overline{\left(\vec{y}_{ensmean}-\vec{y}_{truth}\right)^{2}}}
\end{equation}

With $\vec{y}_{ensmean}$ the mean of the ensemble.
All computations are performed on the regular ERA5 lat-lon grid, without taking the different grid-density towards the poles into account. We evaluate the spread information in two ways. First, we compare the mean spread to the mean error. As mentioned, in a perfect ensemble these should be equivalent. The average error of forecasts of chaotic systems grows with increasing lead time, before eventually saturating at some climatological level. Therefore, the average spread as a function of lead time should follow the average error during the initial error growth phase. Secondly, we compute the correlation between spread and error of all forecasts for a given lead time. Since the spread is only a predictor of the expected, or average, error this correlation
would not be 1 even for a perfect ensemble. However, if the spread contains useful information about the day-to-day uncertainty in the forecasts, the correlation should be significantly larger than zero \cite{buizza_potential_1997-2}.

\subsubsection{CRPS}

In addition to spread-error relations, we also compute the Continuous Ranked Probability Score (CRPS). This metric is widely used for evaluating ensemble forecasts (e.g. \cite{rasp_neural_2018-2}). The CRPS is defined as

\begin{equation}
CRPS(F, y) = \int_x (F(x) - H(x - y))^2 dx
\end{equation}
where $F(x)$ is the cumulative distribution
    function (CDF) of the forecast distribution, $y$ the true value  and $H(x)$  the
    Heaviside step function,

We use the python library "properscoring" which estimates the CRPS via the empirical cumulative distribution function.
We compute the CRPS for each gridpoint of the forecasts separately, and then average over all gridpoints. Since CRPS is a probabilistic measure, in contrast to RMSE, we do not compute CRPS for the unperturbed forecasts.

\section{Results}

Each of the ensemble methods (except the \emph{multitrain} ensemble method) has one or more free parameters: the dropout rate $p_{drop}$ for the drop-ensemble, the initial perturbation scale $\sigma_{rand}$ for the random ensemble, and $\sigma_{svd}$, $T_{svd}$ and $n_{svs}$ for the SVD ensemble.

\begin{figure}
\includegraphics[width=1\textwidth]{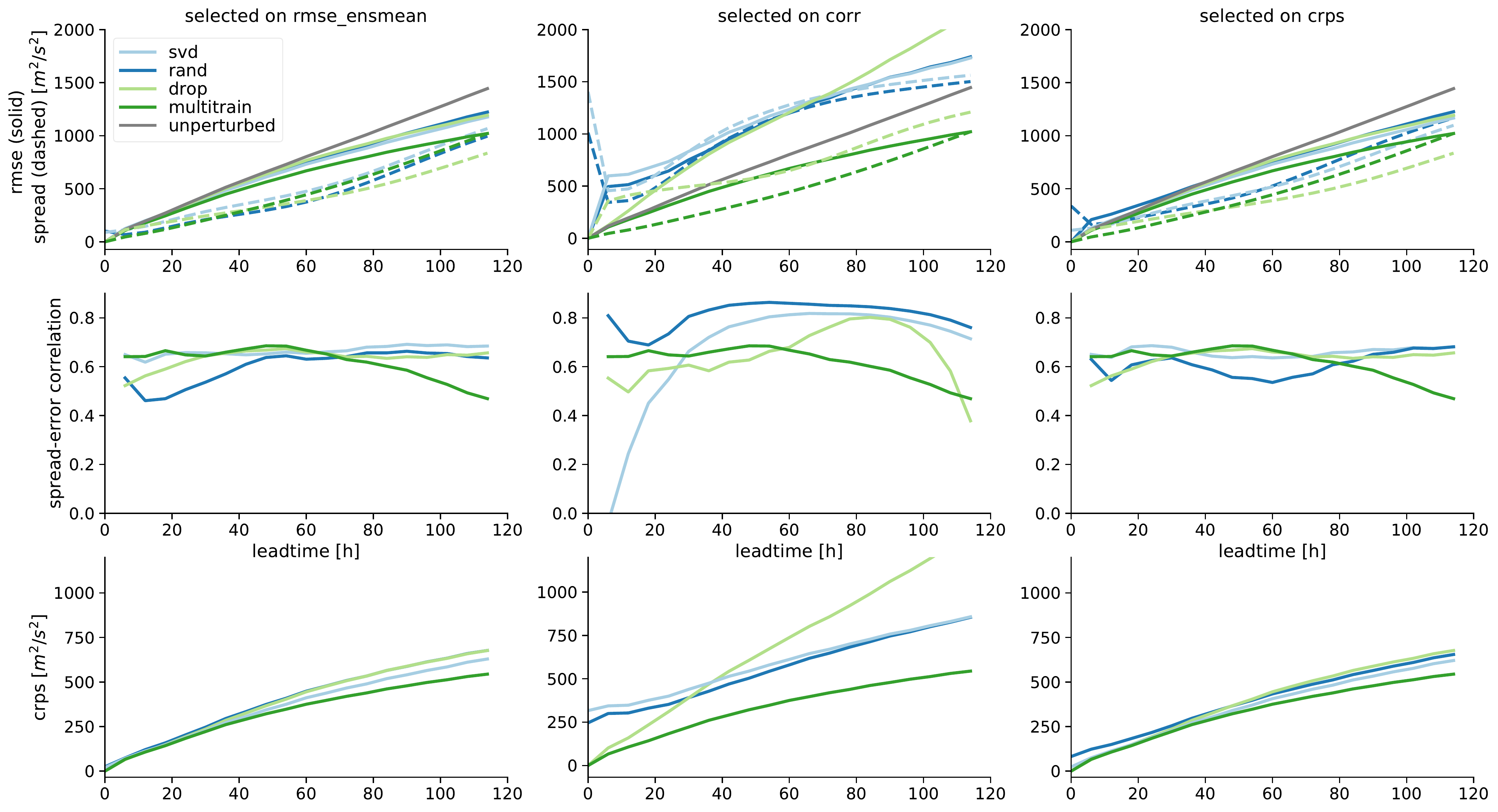}
\caption{\label{fig:main-res} Results for the best parameter settings for all 4 methods. The parameters for each method were selected such that at a lead time of 60h the ensemble-mean RMSE is minimized (column 1), the spread-error correlation is maximized (column 2) or the CRPS is minimized (column 3). In Addition to the ensemble, the RMSE of the unperturbed forecasts is shown (grey). In the uppermost rows, the solid lines show the RMSE, and the dashed lines the ensemble spread.}
\end{figure}

We start by finding the parameter settings that provide the best score at a lead time of 60 hours, separately for three different scores: ensemble-mean RMSE, spread-error correlation and CRPS. In principle, one could also use a weighted combination of these scores to find a single best parameter setting. However, since objectively finding a reasonable weighting is very difficult, we chose not to attempt this. The resulting parameter combinations are shown in table 1. The resulting scores for all methods are shown in fig. \ref{fig:main-res}, with the largest ensemble sizes for each method (20 for \emph{multitrain}, 100 for the other methods). Each column corresponds to a different parameter selection method (minimal RMSE, maximal spread-error correlation and minimal CRPS). The results are relatively similar when selecting on minimal RMSE and minimal CRPS, as also evidenced by the closeness of the resulting parameters (table 1). When selecting on RMSE, the error and CRPS are lowest for the \emph{multitrain} method, with the other three all being similarly higher. The unperturbed single forecasts are poorer than any of the 4 ensembles in RMSE. All 4 methods also have very similar spread (dashed lines), although slightly larger differences emerge when selecting on CRPS rather than RMSE. Spread-error correlation is roughly between 0.45 and 0.65 for all methods and lead times considered. The \emph{multitrain} ensemble correlation decreases slightly at longer leadtimes, while the  \emph{drop} and  \emph{rand} ensembles have lower correlation at shorter lead-times. When selecting on maximum correlation, the resulting parameters are quite different from the RMSE and CRPS selections, with much higher initial perturbation rates (table 1), resulting in much larger spread. While this leads to better correlation for most lead times, both in the  \emph{rand} and the  \emph{svd} ensembles, it comes at the cost of degraded RMSE and CRPS performance.

\begin{figure}
\includegraphics[width=0.5\textwidth]{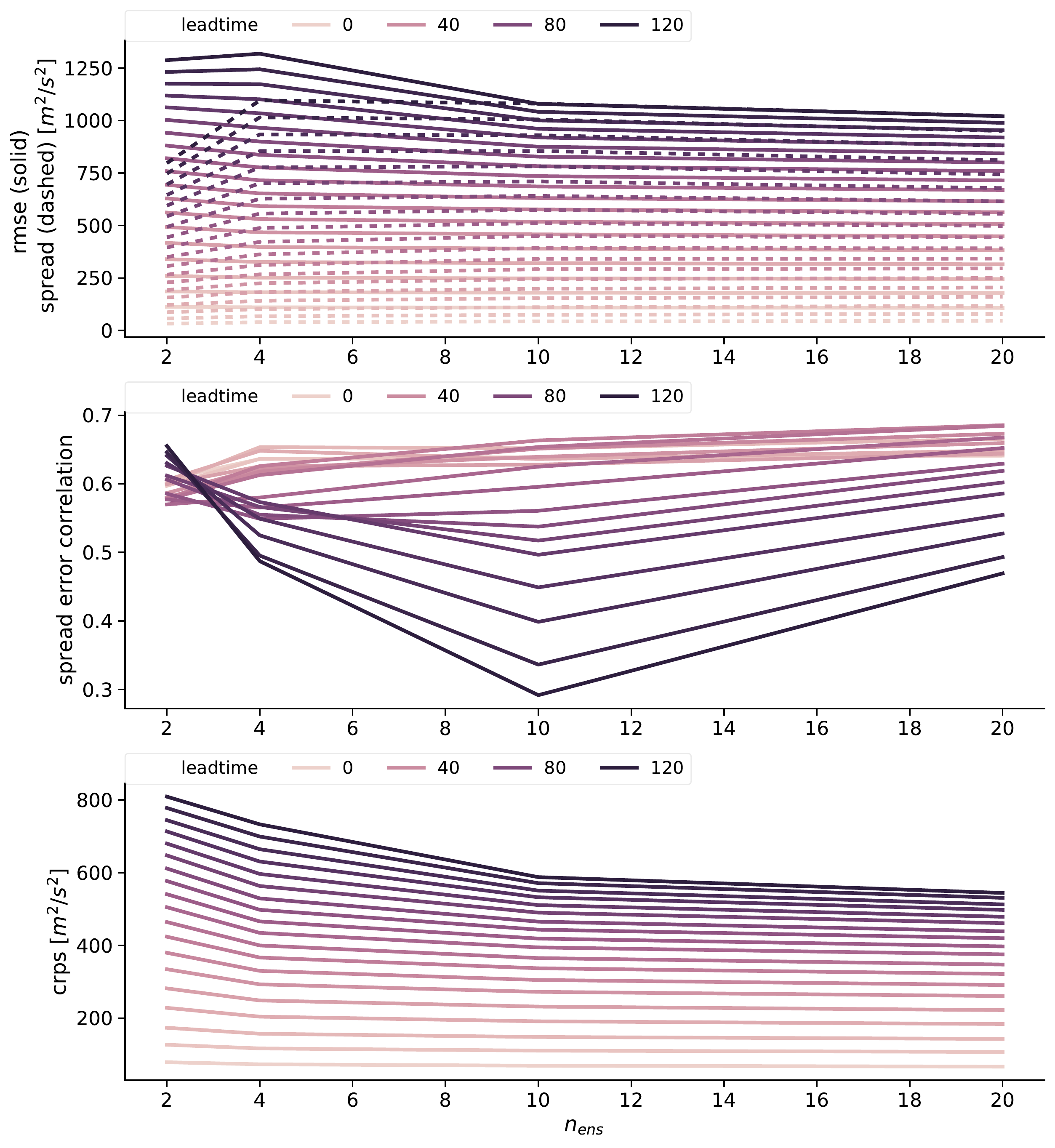}
\caption{\label{fig:netens-sens} Sensitivity of the \emph{multitrain} method to the number of ensemble members $n_{ens}$. The x-axis shows the number of ensemble members, the line colors indicate the lead-time (in 6h intervals). }
\end{figure}

Above, we have discussed the results when using parameters optimised on different targets. We now consider the sensitivity of the results to the parameter choices. For this, we vary a single parameter, while leaving the other parameters fixed to the ones obtained by minimizing RMSE (values shown in table 1). Figure \ref{fig:netens-sens} shows the sensitivity of the \emph{multitrain} method to the number of ensemble members. Each line represents a single lead time. For a very small ensemble with only 2 members, RMSE is higher than for the full ensemble, and the spread is too low. With increasing ensemble size, RMSE decreases and the spread grows to match RMSE. CRPS continuously decreases with ensemble size. The spread-error correlation continuously increases with ensemble size for short lead-times. For longer lead-times, it first decreases, and then increases again. While this seems counter-intuitive, we hypothesize that it is an artefact of the higher RMSE of the very small ensembles. As we have seen above, poor forecasts (large RMSE) can nonetheless display a higher spread-error correlation than more skillful forecasts. The very small ensembles have a high RMSE, yet it appears that this error is well-predicted by the ensemble spread.

\begin{figure}
\includegraphics[width=1\textwidth]{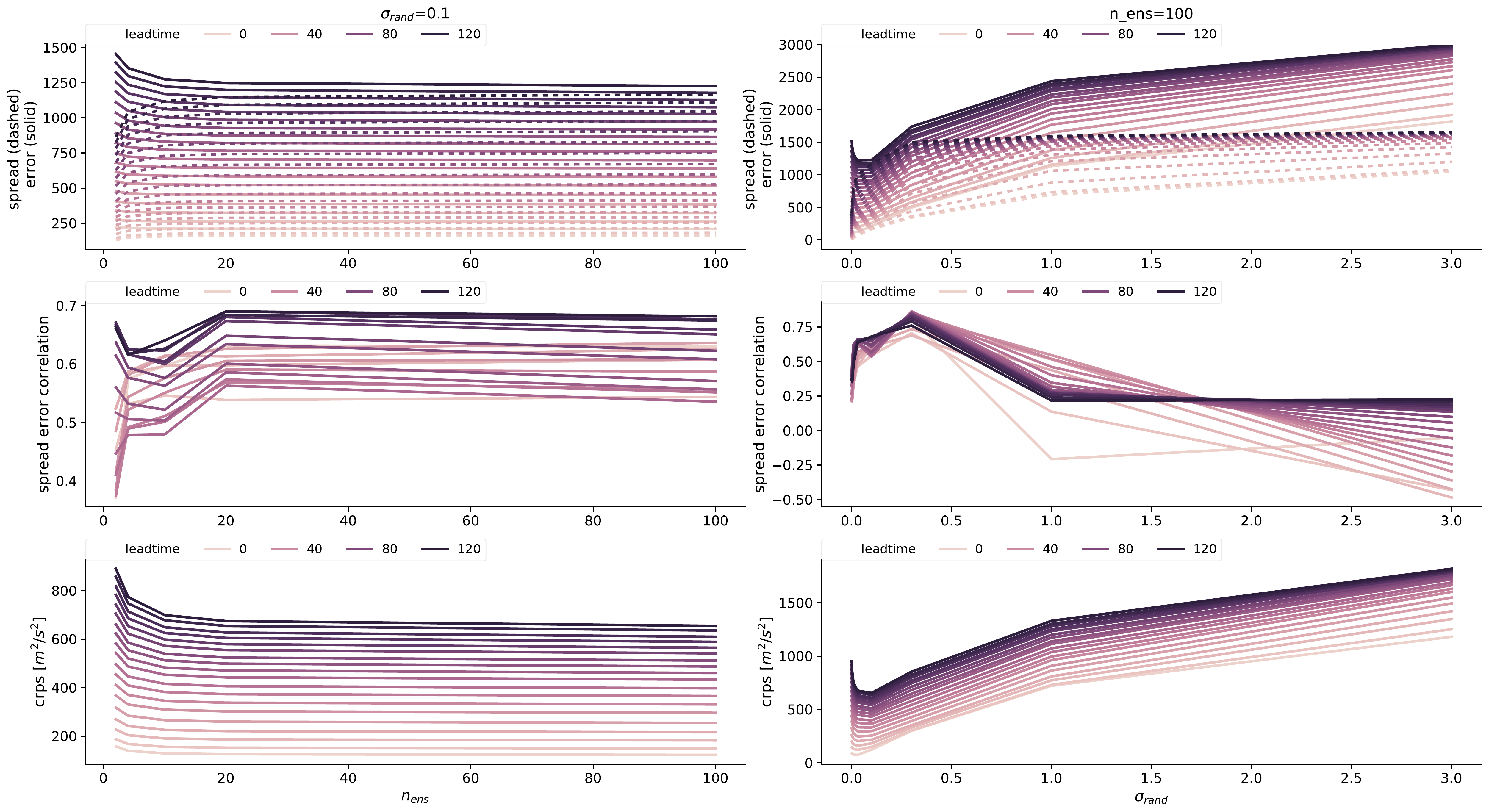}
\caption{\label{fig:rand-sens} Sensitivity of the  \emph{rand} ensemble method to the number of ensemble members $n_{ens}$ (left) and $\sigma_{rand}$ with $n_{ens}=100$ (right). }
\end{figure}

The sensitivity of the  \emph{rand} ensemble is shown in fig. \ref{fig:rand-sens}. The left and right panels show the sensitivity to ensemble size and $\sigma_{rand}$, respectively. The sensitivity to ensemble size is similar to the \emph{multitrain} method. For the initial perturbation scale, there are 2 different optimums: one for CRPS and RMSE at smaller values, and one for spread-error correlation at larger values. This mirrors the values shown in table 1. As expected, we also find an increasing error with increasing initial perturbation. Spread, on the other hand, saturates with increasing $\sigma_{rand}$.

\begin{figure}
\includegraphics[width=1\textwidth]{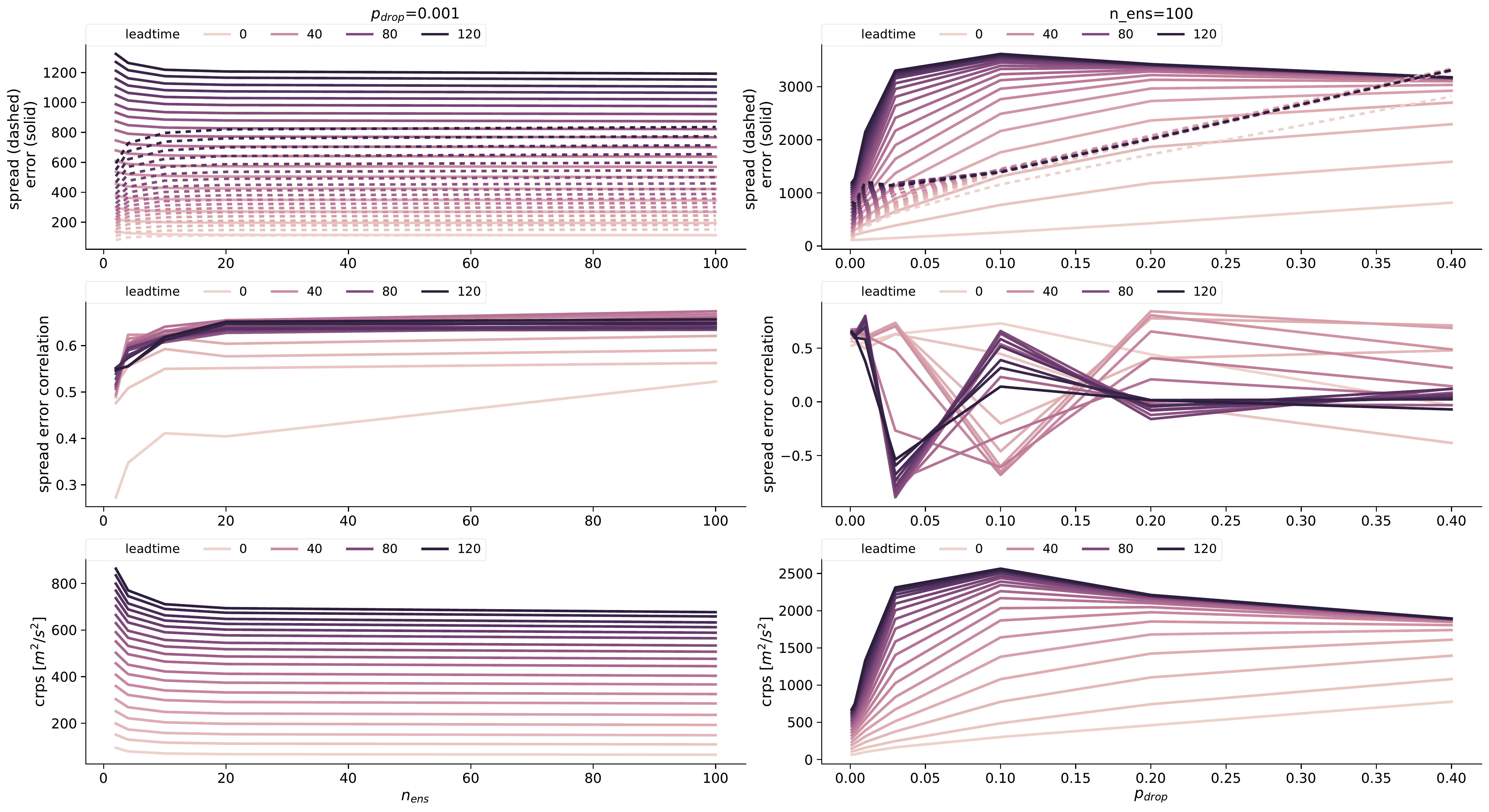}
\caption{\label{fig:drop-sens} Sensitivity of the dropout ensemble method to the number of ensemble members $n_{ens}$ (left) and $p_{drop}$ with $n_{ens}=100$ (right).}
\end{figure}

Results for the \emph{drop} ensemble are shown in fig. \ref{fig:drop-sens}. In contrast to the \emph{rand} ensemble, spread increases with increasing $p_{drop}$, while RMSE saturates.

\begin{figure}
\includegraphics[width=1\textwidth]{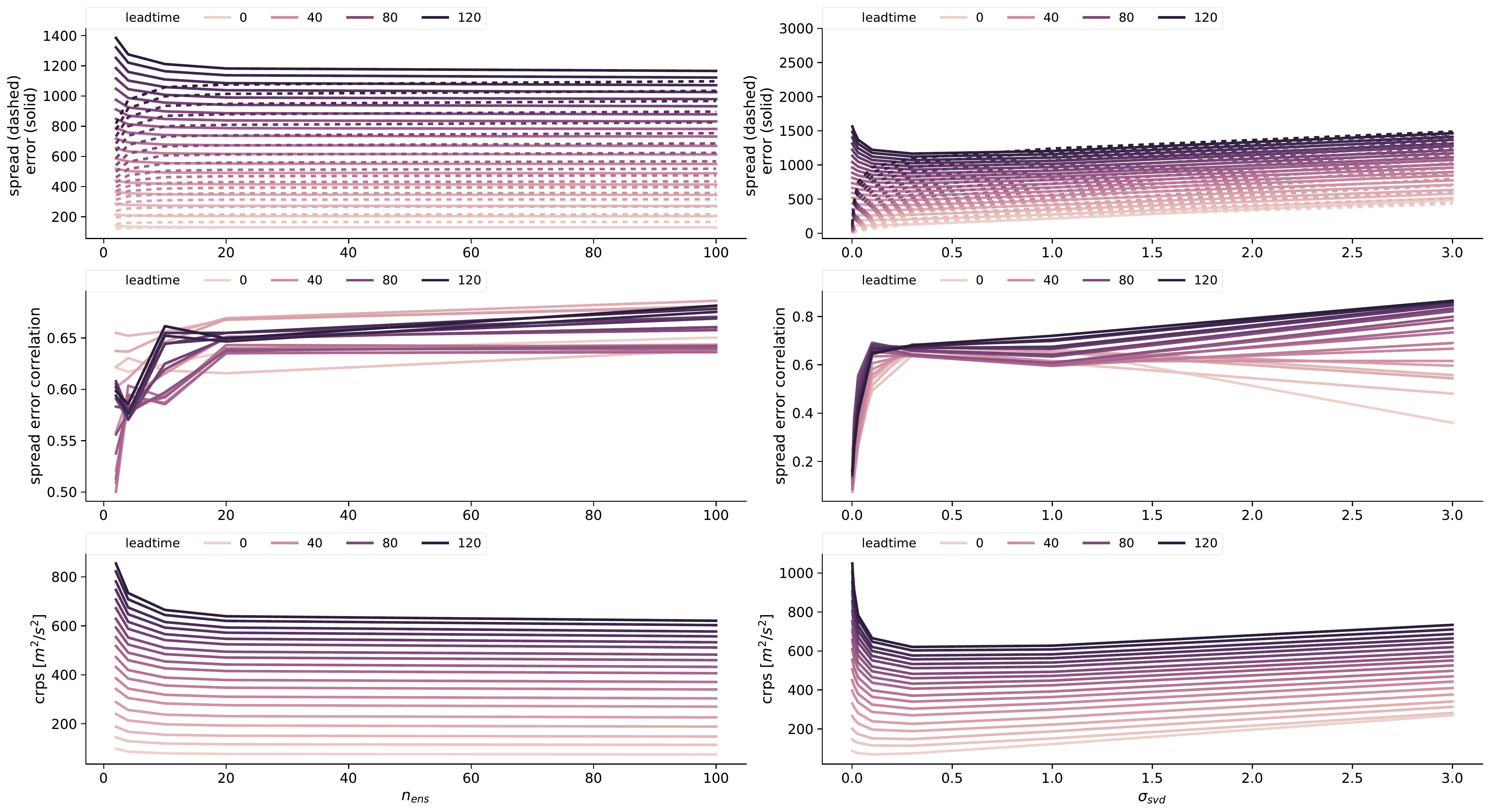}
\caption{\label{fig:svd-sens} Sensitivity of the \emph{svd} ensemble method to the number of ensemble members $n_{ens}$ (left) and $\sigma_{svd}$ (right), with the remaining parameters fixed at the optimal ones obtained when optimizing RMSE and $n_{ens}=100$.}
\end{figure}

\begin{figure}
\includegraphics[width=1\textwidth]{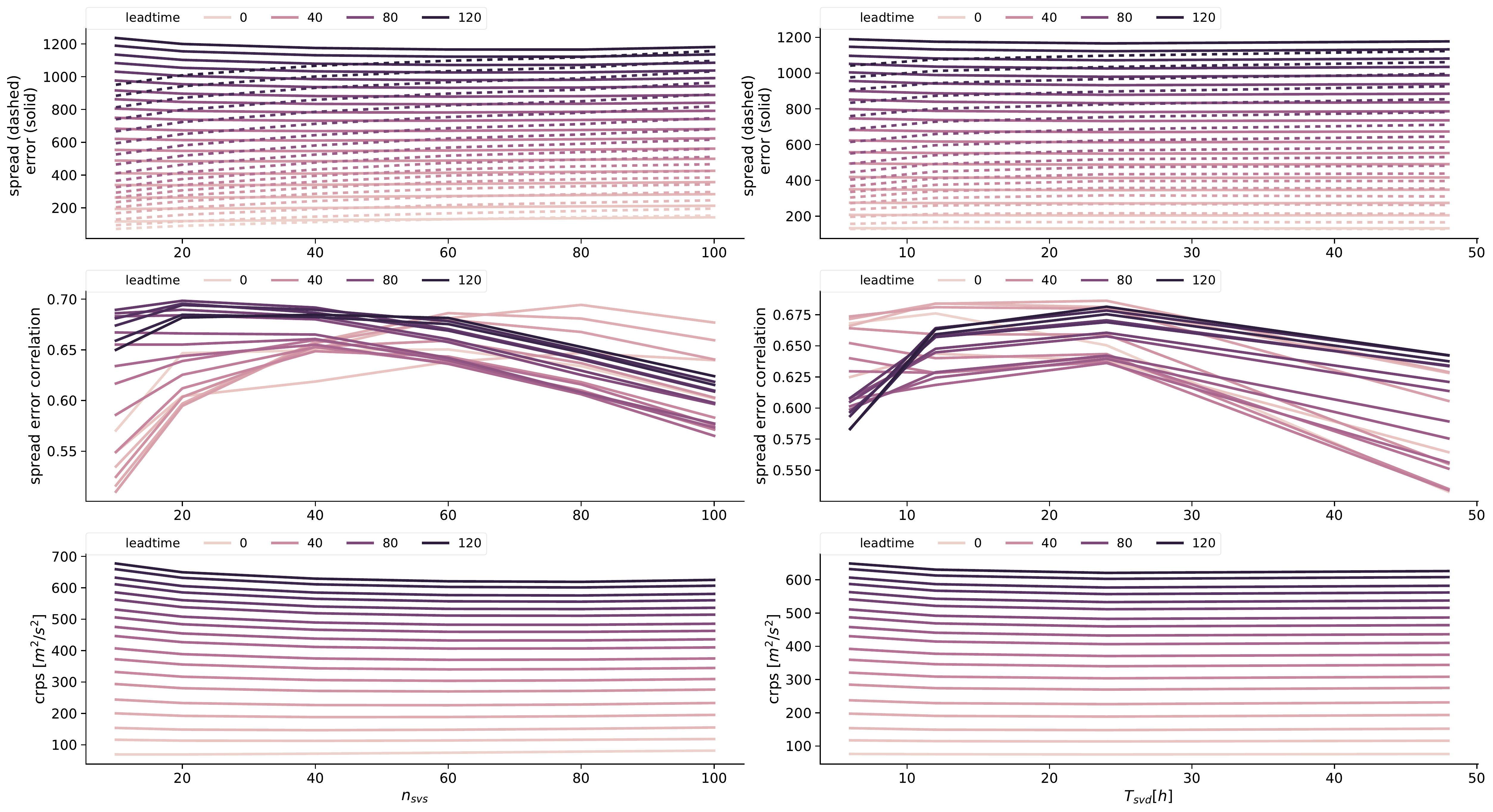}
\caption{\label{fig:svd-sens2} Sensitivity of the \emph{svd} ensemble method to $n_{svs}$ (left) and $T_{svd}$ (right), with the remaining parameters fixed at the optimal ones obtained when optimizing RMSE.}
\end{figure}

Finally, the results for the SVD ensemble, which has the largest number of parameters, are shown in fig. \ref{fig:svd-sens} and \ref{fig:svd-sens2}. Figure \ref{fig:svd-sens} shows the sensitivity to $n_{ens}$ and $\sigma_{pert}$. As with the other methods, RMSE and CRPS decrease with increasing ensemble size. 
Correlation increases for short lead-times, while at longer lead times the same behaviour as for \emph{multitrain} and \emph{rand} is visible. When going from very small to slightly larger ensemble sizes, the correlations drops. It then increases again for even larger ensemble sizes. Again, we hypothesise that this may be caused by the higher RMSE of the very small ensembles. Regarding sensitivity to $\sigma_{pert}$, there is a clear optimum at 0.3 for minimizing RMSE and CRPS, whereas higher values lead to better spread-error correlation, at the cost of higher RMSE.
Figure \ref{fig:svd-sens2} shows sensitivity to $n_{svs}$ and $T_{svd}$. The number of leading singular vectors $n_{svs}$ only has a minor influence on RMSE and CRPS (a small increase in skill with increasing $n_{svs}$), while it affects more distinctly the spread-error correlation. Here, there is a clear optimum for intermediate $n_{svs}$ values, with the exact number being different for each lead time. Specifically, the longer the lead time, the lower the optimal $n_{svs}$. The lead time over which the SVD is performed ($T_{svd}$) has hardly any impact on RMSE, and only small influence on CRPS (a small decrease with increasing $T_{svd}$). Just as $n_{svs}$, it has a more profound influence on the spread-error correlation. However, contrary to $n_{svs}$ the optimal value is only weakly dependent on lead time. For most lead times it is $T_{svd} = 24h$, but is smaller for some short lead times.

\begin{figure}
\includegraphics[width=1\textwidth]{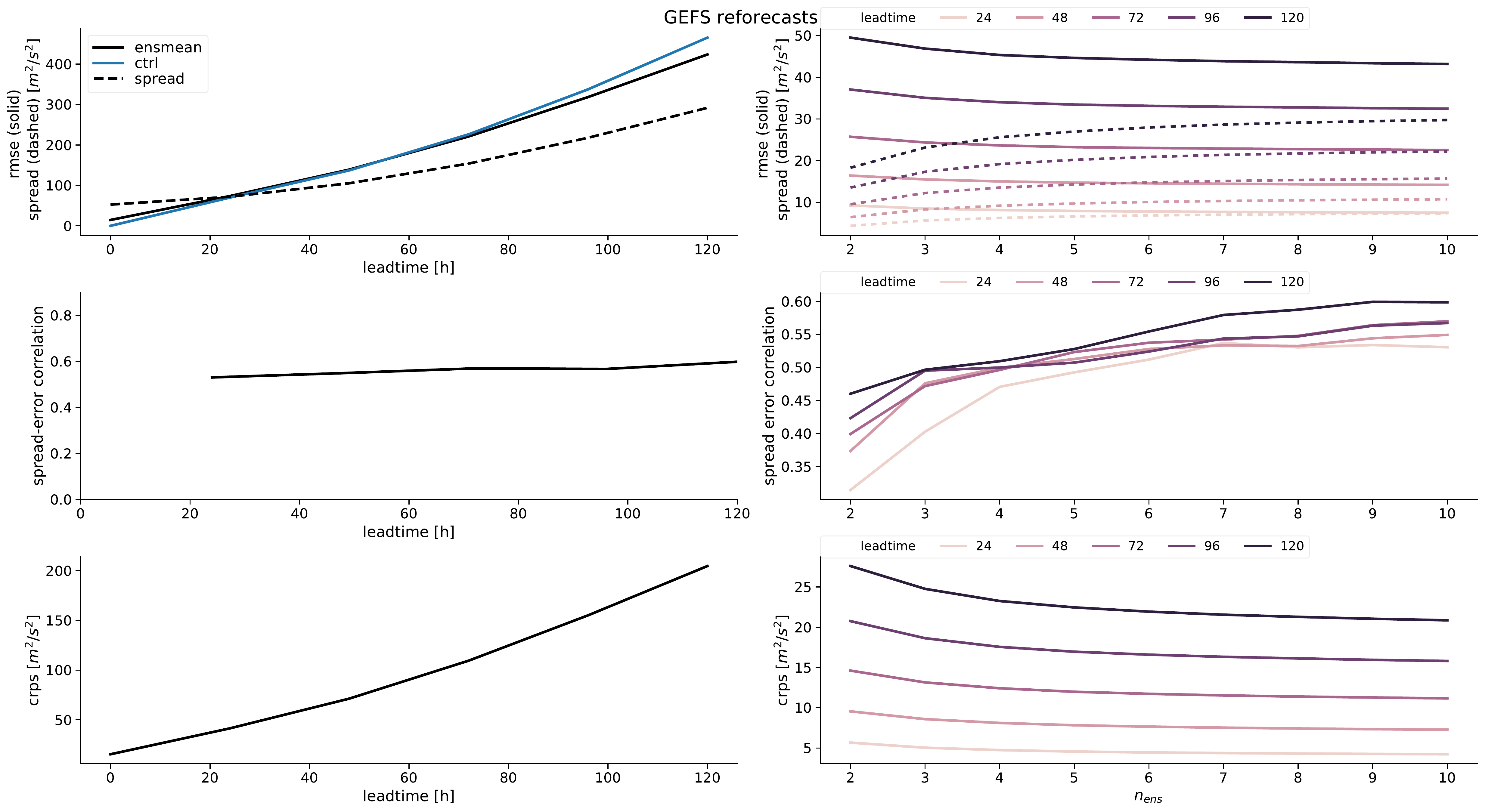}
\caption{\label{fig:gefs} Skill of the GEFS ensemble reforecast (left) and sensitivity of the scores to the number of ensemble members $n_{ens}$ (right), evaluated on a 1\degree $\times$ 1\degree grid.}
\end{figure}

In order to put the above results into context, we also consider the skill of the ensemble forecasts from the GEFS reforecast dataset, including the sensitivity to the ensemble size (fig. \ref{fig:gefs}). RMSE and CRPS are much lower than for the neural network forecasts. Spread-error correlation, on the other hand, is comparable. Regarding sensitivity to ensemble size, RMSE and CRPS decrease monotonically with increasing size, whereas the spread-error correlation increases.

\begin{table}
    \caption{Best parameters for each selection metric.}
\begin{tabular}{lrrrrr}
  selection metric &  $\sigma_{rand}$ &  $\sigma_{svd}$ &  $n_{svs}$ &  $T_{svd} [h]$  &  $p_{drop}$ \\\hline
 RMSE &             0.03 &             0.3 &       40 &       24 &       0.001 \\
         corr &             0.30 &             3.0 &      100 &       48 &       0.010 \\
         CRPS &             0.10 &             0.3 &       60 &       24 &       0.001 \\
\end{tabular}
\end{table}

\section{Discussion and conclusions}

In this paper, we have presented and tested four methods for transforming a deterministic neural network weather forecasting system into an ensemble forecasting system. Two of these methods perturb initial conditions (one with random perturbations, one with perturbations based on the SVD technique). The third method retrains the neural network, creating a slightly different neural network each time, and the fourth methods uses dropout in the network to generate an ensemble.

These methods were only partly borrowed from NWP development. In the latter, it is common to strictly differentiate between perturbation of the initial fields (which can be done, for example, with the SVD technique but also with other methods such as bred-vectors), and model perturbations. The reason for this partition are the different sources of uncertainty: there is uncertainty in the initial fields, and there is uncertainty in the models themselves, because they are not perfect. In principle, this also translates to neural network forecasts (were both the initial field and the trained neural network are not perfect). However, the network itself is dependent on errors in the initialisation data, as we train the network on the same dataset that we then use to initialise the forecasts (although we naturally use two different time periods for the training and the testing). The distinction between the two sources of uncertainty is therefore not as clear-cut as in conventional NWP. It would nonetheless be possible, just as in NWP, to combine perturbations of the initial fields with perturbations of the networks -- something which we have not tested here. More generally, neural networks are a very results-oriented tool, in that they do not necessarily attempt to model the processes underlying the evolution of a given system, but only to optimise a specified output. Whether the only way for them to make skillful forecasts is to approximate the underlying processes as well as possible is a question which we do not attempt to answer here.

Based on the above, we chose to compare all of our different ensemble methods to one another. For many (albeit not all) users, it will not matter how the ensemble is generated, as long as it has (probabilistic) skill. At the same time, we recognize that some skilled users may tailor their interpretation of the ensemble forecasts to the method the ensemble is generated with, and may find the machine-learning approaches described here unsuitable for their purposes.

All ensembles were evaluated by analyzing Root Mean Square Error of the ensemble mean forecast, ensemble spread and CRPS, and compared to a NWP model. The neural network architecture we used has previously been used in the literature for performing unperturbed (or "deterministic") weather forecasts. Each of the ensemble methods creates ensembles whose mean improves over the unperturbed neural network forecasts, with the method that retrains the network achieving the highest improvement both in ensemble-mean RMSE and CRPS. All methods have relatively similar spread-error correlation, with random initial perturbations and dropout performing slightly worse than the other methods, except at long lead times (beyond ~3 days) where the \emph{multitrain} ensemble displays rapidly decreasing correlations. As a caveat, spread-error correlation is a somewhat disputed measure (e.g. \citeA{10.1175/MWR-D-12-00111.1}), and should not be over-interpreted. 
Except for the network retraining, the methods have free parameters that need to be chosen. We found that optimizing them on ensemble mean RMSE and CRPS leads to relatively similar results. Optimizing on spread-error correlation turned out to be problematic. While it does lead to higher spread-error correlations than when optimising on RMSE and CRPS, this came at the cost of a markedly degraded performance in the latter metrics. This may be linked to the tendency of ensemble forecasts to display the highest spread-error relationships for forecasts with unusually large (or small) spread (e.g. \cite{grimit2007measuring} and references therein). All ensemble network forecasts are outperformed by NWP forecasts from the GEFS reforecast dataset in both RMSE and CRPS. This is unsurprising, given the low skill of the deterministic network architecture \cite{weyn_can_2019}. In terms of spread-error correlation, the neural network ensembles have a performance comparable to the NWP forecasts. 

An important caveat of our results is that the errors of the network forecasts do not show exponential growth with increasing lead time, in contrast to NWP models. This might have implications for the theoretical grounding of ensemble techniques, and especially the SVD technique originally developed for NWP models, in our analysis. The fact that the neural networks here do not show exponential growth (fig. C1) is indeed a warning sign that they do not actually model the underlying system, which is known to be chaotic and thus must show exponential error growth in at least one dimension. Specifically, it is not clear that the insights obtained here may be directly applicable to a hypothetical future  neural network system with high forecast skill and exponential error growth, such as the method recently proposed by \citeA{rasp2020purely}. Additionally, an ensemble forecasting system whose statistics do not match the expected behavior of the dynamical system it is attempting to model, can lead to the situation where the statistics of the ensemble forecasts accurately model the forecast error, but are quite far away from the real dynamical system. Whether this makes such a forecast meaningless, or whether the ensemble statistics nonetheless provide valuable information, can probably only be answered on an application by application basis.

The good performance of the \emph{multitrain} ensemble points to an important side-result, namely that retraining the network gives different forecasts. While this is desirable for exploring the space of possible future states -- as is wished in ensemble forecasting -- it also has implications for deterministic forecasting. The literature to date has focused on deterministic neural network weather forecasts, and our results show that the uncertainty derived from network training is a potentially important aspect in this context. To our best knowledge, this has not been shown before. Whether the fact that retraining the network gives different forecasts each time is an intrinsic property of neural network forecasts of chaotic systems, or whether this is a limitation in our current architecture, remains open. Specifically, this behavior may indicate that the networks attain (different) local minima of the loss function, as opposed to a global minimum solution. From a NWP development perspective, one can argue that our results imply that the forecasts are more sensitive to the model itself than to the initial conditions, which is in contrast to state-of-the art NWP systems. Finally, in operational practice forecasters would need to be aware that after retraining the neural network model, the performance of an older training realization for a particular case-study weather event would not necessarily be representative, even though the skill averaged over all forecasts would be nearly unchanged.

While this study focused on weather prediction, the principles presented here can also be applied to the forecasting of other initial value problems. Indeed, the SVD technique can in principle be used with any end-to-end differentiable function. Therefore, it could also be used for hybrid numerical and neural network models, as long as they are differentiable. The same holds for the other three methods. SVD itself is also differentiable, making it possible to include the generation
of perturbed initial states in the neural network training procedure itself. In this way, one could for example optimize both on ensemble mean error and ensemble mean spread at the same time, or on  CRPS, as in \citeA{gronquist2020deep}. Furthermore, applying the SVD technique to neural networks is in fact easier than for numerical models, as the latter require making a tangent linear version of the model first, either through re-coding the model, or with automatic differentiation techniques. The computation of the singular vectors could also be sped up with the Lanczos algorithm, which is faster than explicitly computing the Jacobian first. Finally, the fact that we could directly apply a method developed in the context of NWP models to neural networks shows that there are potential synergies between these two forecasting concepts, notwithstanding the many differences discussed above. Indeed, more concepts developed for NWP, beyond the SVD technique, may be transferable to machine learning based weather forecasts.

The original aim of this study was to provide a proof-of-concept for performing neural network-based ensemble weather forecasts. Our results confirm previous results that significant improvements in forecast skill need to be made before neural network forecasts may compete with NWP models. At the same time, we show that existing network architectures can already be used to provide probabilistic forecasts with uncertainty estimates comparable to those of NWP models.

\subsection*{Code and data availability}The software used for this study was developed in python, using the tensorflow framework, and is available in the repository (\url{
https://doi.org/10.5281/zenodo.4013698}) and on S. Scher's github page (\url{https://github.com/sipposip/ensemble-neural-network-weather-forecasts}). The data underlying the figures is also available in the repository. ERA5 data can be freely obtained through the Copernicus Climate Change Service at: \url{https://cds.climate.copernicus.eu/cdsapp\#!/dataset/reanalysis-era5-pressure-levels?tab=overview}. The GEFS reforecast data can be freely downloaded from \url{https://psl.noaa.gov/forecasts/reforecast2/download.html}.

\appendix
\section{}

\begin{algorithm}[H]
\begin{algorithmic}

\STATE require $N_{ens}$
\STATE require $N_{svs}$
\STATE require $\sigma_{pert}$
\STATE $J\gets jacobian(NN(\vec{x}_{init}))$
\STATE $S\gets SVD(J)$
\STATE $S\gets S[1..N_{svs}]$
\FOR{i in $N_{ens}/2$}
\STATE $\vec{p}=\vec{0}$
\FOR{j in $N_{svs}$}
\STATE $r\gets normal(0,1)$
\STATE $r\gets clamp(r,-3,3)$
\STATE\texttt{$\vec{p}\gets\vec{p}+r\cdot S_{j}\cdot \sigma_{pert}$}
\ENDFOR
\STATE $\vec{x}_{init,pert1}'\gets\vec{x}_{init}+\vec{p} $
\STATE $\vec{x}_{init,pert2}'\gets\vec{x}_{init}-\vec{p} $
\ENDFOR
\end{algorithmic}\caption{\label{alg:computation-of-the}computation of initial perturbations
$x_{init}'\left(\ensuremath{\vec{x}_{init}},\ensuremath{NN}\right)$.}
\end{algorithm}

\section{TLM-test}
To check the validity of the Tangent Linear Model (TLM) derived through the computation of the Jacobian of the neural network forecasting system, we perform a basic TLM test. We specifically compare the perturbations of a forecast made with the TLM model to the perturbation of the forecast of the NN model itself.

If we have an initial perturbation with pattern $\vec{x}'$ and scale $\sigma$, the perturbation obtained with the TLM is
\begin{equation}
\vec{y}'_{tlm}=J\cdot\left(\sigma\vec{x}'\right)
\end{equation}
and the perturbation obtained with running the actual NN forecast system is
\begin{equation}
\vec{y}'_{NN}=NN\left(\vec{x}+\sigma\vec{x}'\right)-NN\left(\vec{x}\right)
\end{equation}

For our test, we use the leading singular vector for $x'$. Then, for each initial condition, we compute the area mean of $\vec{y}'$ for each of the two methods for different values of $\sigma$. The results are shown in fig. B1. As can be seen, for small $\sigma$, the TLM response follows reasonably closely the actual response of the NN system. This supports the validity of the TLM as a reasonable approximation for small perturbations.

\begin{figure}
\includegraphics[width=1\textwidth]{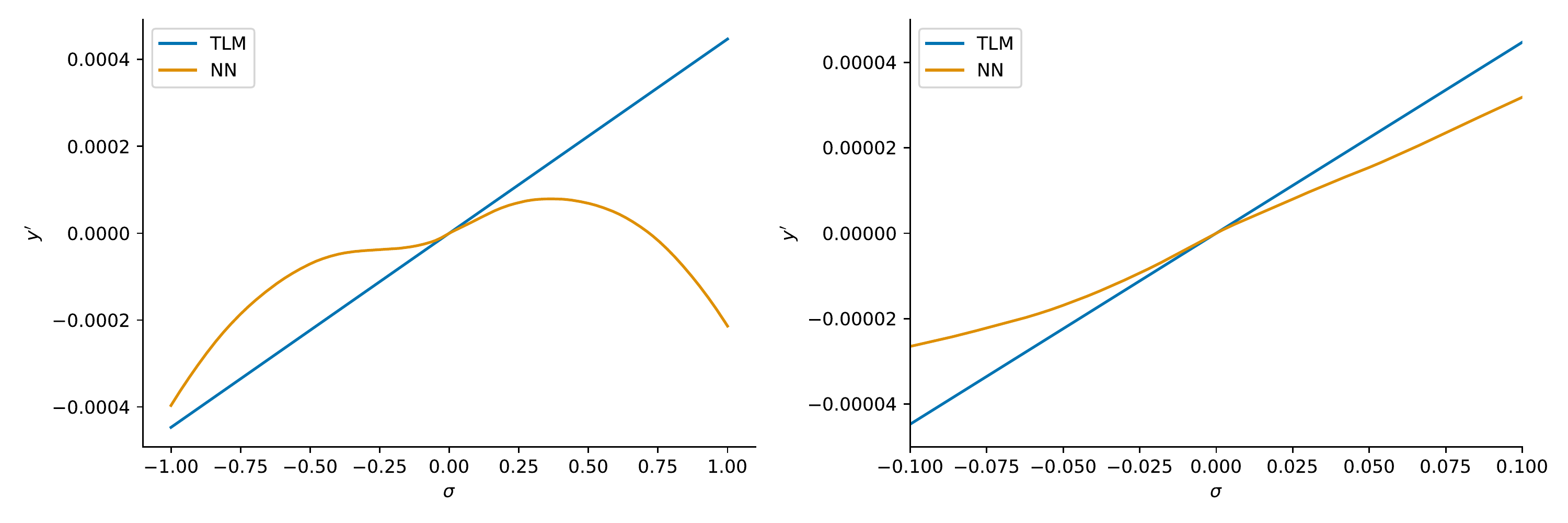}
\caption{Results of the Tangent Linear Model (TLM) test. The blue line shows the mean response of the TLM model derived from the NN forecast sytem to perturbations, and the yellow line shows the response of the NN system itself. The right panel is a closeup of the left panel.}
\end{figure}
\section{}
\begin{figure}
\includegraphics[width=1\textwidth]{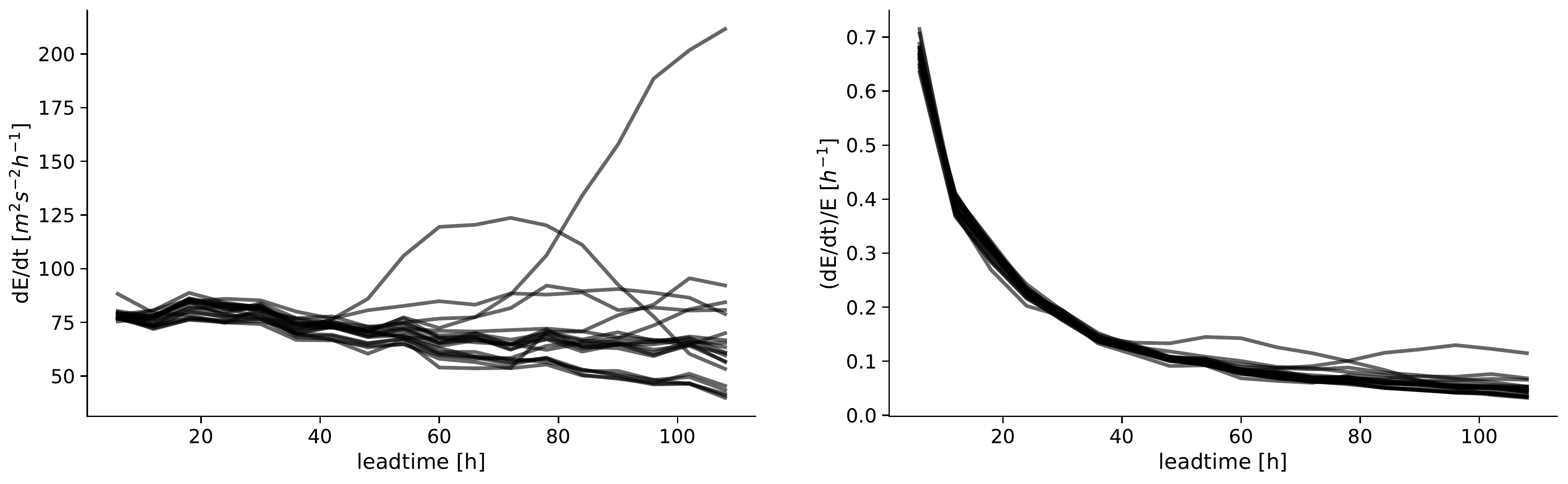}
\caption{Error growth rate over time. Left panel: absolute error growth (dE/dt), right panel: relative error growth (dE/dt)/E.}
\end{figure}

\subsection*{Author contributions}SS conceived and implemented the methods presented in the study and drafted the manuscript. Both authors designed the study, interpreted the results and improved the manuscript.

\acknowledgments
S.S. was funded by the Dept. of Meteorology of Stockholm University. G.M. was partly supported by the Swedish Research Council Vetenskapsr\aa det (grant no.: 2016-03724).  The computations were performed on resources provided by the Swedish National Infrastructure for Computing (SNIC) at the High Performance Computing Center North (HPC2N) and National Supercomputer Centre (NSC) partially funded by the Swedish Research Council through grant agreement no. 2018-05973.
We would like to thank Peter Dueben and one anonymous reviewer for their input which proved essential for the correct framing and discussion of our results.


%
%

\bibliography{nn_ensemble_nwp}

%
%
%
%
%

\end{document}